\documentclass{ckm}                 % twocolumn proceedings style
\usepackage{txfonts}            % if you have it, to get Times family
\confname{Workshop on the CKM Unitarity Triangle, IPPP Durham, April
  2003}

\title{Radiative Penguin decays}

\author{S.Playfer}

\address{School of Physics, University of Edinburgh}

\begin{document}

\begin{abstract} A review of recent experimental results on 
radiative Penguin decays, and their interpretation within 
the Standard Model. 
\end{abstract}

%% \maketitle needs to be after the author and address info and the
%% abstract... 
\maketitle

%% standard LaTeX from here on...

\section{Introduction}

Penguin decays are flavour-changing neutral current (FCNC)
processes, described primarily by second order weak interactions, 
where the decay occurs through the emission and reabsorption 
of $W$ bosons. 
 
\begin{figure}
\hbox to\hsize{\hss
\includegraphics[width=0.6\hsize]{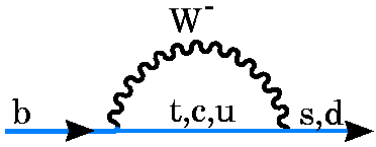}
\hss}
\caption{Penguin diagram for the $b\to s(d)$ transition}
\label{fig:penguin}
\end{figure}

In this review I will focus on the recent experimental results 
on the transition $b\to s$ coming from radiative penguin 
decay measurements by the B-factory experiments, 
BaBar, BELLE and CLEO.  The penguin transition has a dominant role in 
these decays, and a significant role in rare 
hadronic B decays, where it complicates the extraction of 
information about the CKM unitarity triangle because of interference with 
spectator diagrams~\cite{penguinpollution}. 
Measurements of the radiative decays can be used to 
constrain calculations of the penguin 
contributions to the rare hadronic decays.

The measurement of the gamma energy spectrum in the
process $b\to s\gamma$ has recently been used by CLEO 
to improve the extraction of the CKM elements $V_{cb}$ and $V_{ub}$ from 
inclusive semileptonic $b$ decays~\cite{CLEOshape}. 
Details of these moments analyses can be found in~\cite{Luke}

I will also discuss the FCNC transition $b\to d$, which we expect to find
experimental evidence for in the near future. Measurements of 
this transition are sensitive to the CKM element $V_{td}$.  

For completeness, I note that the FCNC transition $s\to d$ has a role in 
the interpretation of $\epsilon '/\epsilon$ in the 
neutral Kaon system. 
The Standard Model calculation of 
$\epsilon '/\epsilon$ has large uncertainties which depend on 
the interference between penguin contributions~\cite{epsilonprime}. 
I also note, that the FCNC transition $c\to u$ 
is heavily suppressed by CKM factors, and because the $b$ quark 
is much lighter than the $t$ quark.

\section{Measurements of $B\to K^*\gamma$}

The exclusive decay $B\to K^*\gamma$ was first observed by CLEO in 1992~\cite{CLEOkstar}. 
BaBar~\cite{BaBarkstar} and BELLE~\cite{BELLEkstar} have now collected large 
samples of these decays with clean signals on low backgrounds (Figure 2). 
The measured branching fractions and world average are summarised in Table 1.
A recent theoretical prediction for this branching fraction is  
$(7.1\pm 2.5)\times 10^{-5}$~\cite{Buchalla}, which is larger than the experimental result, 
but with a 30\% uncertainty due to the $B\to K^*$ form factor.  

\begin{figure}[hbtp]
 \mbox{\scalebox{0.6}{\includegraphics*[100pt,450pt][700pt,750pt]{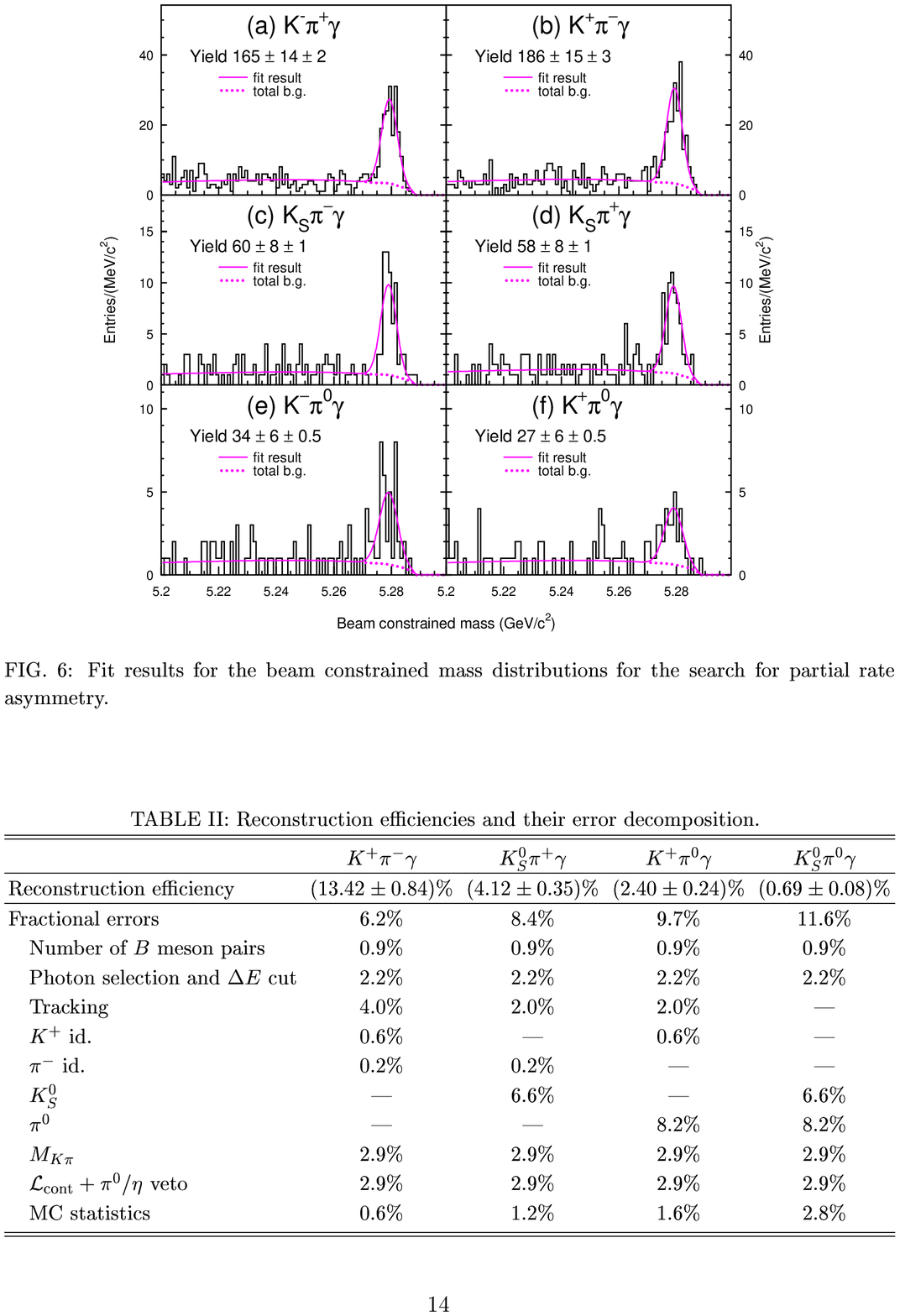}}}
\caption{BELLE results on exclusive $B\to K^*\gamma$}
\label{fig:kstar}
\end{figure}

\begin{table}[htbp]
\begin{center}
\begin{tabular}{|l|c|c|} \hline 
          & $B^0\to K^{*0}\gamma /10^{-5}$ & $B^+\to K^{*+}\gamma /10^{-5}$ \\ \hline 
BABAR 20fb$^{-1}$ &$4.23\pm 0.40\pm 0.22$&$3.83\pm 0.62\pm 0.22$\\ 
BELLE \hspace{0.03cm} 60fb$^{-1}$ &$3.91\pm 0.23\pm 0.35$&$4.21\pm 0.35\pm 0.31$\\
CLEO \hspace{0.25cm} 9 fb$^{-1}$ &$4.55\pm 0.70\pm 0.34$&$3.76\pm 0.86\pm 0.28$\\ \hline
Average   & $4.17\pm 0.20\pm 0.18$ & $3.98\pm 0.28\pm 0.16$\\ \hline
\end{tabular}
\end{center}
\caption{$B\to K^*\gamma$ branching fraction measurements}
\end{table}

There appears to be little prospect of reducing the theoretical uncertainty, so the 
experimental interest is mostly in the measurement of the CP asymmetry. 
This is expected to be $\approx$1\% in the Standard Model, but could be as large 
as 10\% if there are new physics contributions~\cite{CPkstar}. 
Note that the measurements of CP asymmetries in the gluonic penguin decay 
$B\to\phi K_s$ are not very consistent with $B\to \psi K_s$.
Any new physics introduced to explain this discrepancy is likely to 
also appear in radiative penguin decays.   
Table 2 summarizes the measurements of the CP asymmetry in $B\to K^*\gamma$. 
These are consistent with zero and statistics limited. 
With more data it should be possible to reach the 
an accuracy of 1\%. 

\begin{table}[hbtp]
\begin{center} 
\begin{tabular}{|l|l|} \hline
BABAR   & $-0.044\pm0.067\pm 0.012$\\ 
BELLE   & $-0.022\pm 0.048\pm 0.017$\\
CLEO    &  $+0.08\pm 0.13\pm 0.03$ \\ \hline
Combined & $-0.027\pm 0.034\pm 0.015$\\ \hline
\end{tabular}
\end{center}
\caption{CP asymmetry measurements in $B\to K^*\gamma$}
\end{table}

There are also predictions of a few \% for the isospin asymmetry in $B\to K^*\gamma$
~\cite{isokstar}. The experimental situation can be seen from Table 1, 
but note that this Table assumes equal production of $B^+$ and $B^0$ at 
the $\Upsilon$(4S). This assumption introduces an additional uncertainty 
into the isospin asymmetry measurement. 

\section{Measurements of $b\to s\gamma$}

\begin{figure}[hbtp]
\hbox to\hsize{\hss
\includegraphics[width=0.8\hsize]{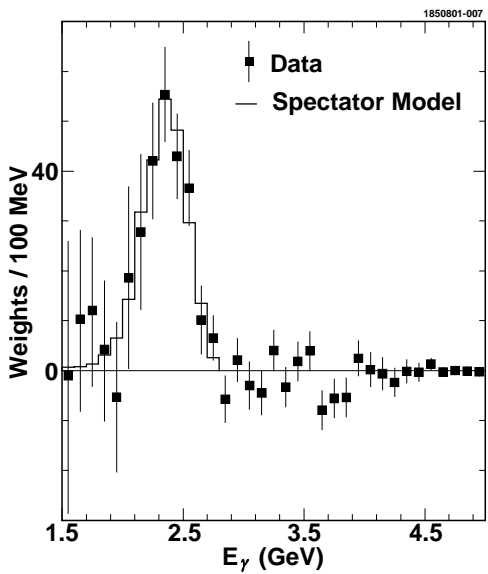}
\hss}
\caption{CLEO measurement of inclusive $b\to s\gamma$}
\label{fig:inclusive}
\end{figure}

The inclusive rate for $b\to s\gamma$ can be quite accurately predicted 
to be $(3.60\pm 0.30)\times 10^{-4}$~\cite{Misiak}. 
The experimental measurements of this inclusive rate provide an 
important constraint on new physics contributions to penguin 
diagrams~\cite{newphysics}.  

Two approaches have been used to measure the inclusive rate. 
The ``fully inclusive'' method measures the high energy photon 
spectrum without identifying the specific $B\to X_s\gamma$ decay modes.
Continuum backgrounds are suppressed with event shape information, 
and then subtracted using off-resonance data.  
B decay backgrounds are subtracted using a generic Monte Carlo 
prediction, which is cross-checked with a $b\to s\pi^0$ analysis.
 
CLEO~\cite{CLEObsg} has published a measurement of the photon 
energy spectrum down to a threshold $E_{\gamma}^* > 2.0$GeV, 
where $E_{\gamma}^*$ is measured in the $\Upsilon$(4S) rest frame
(Figure 3).
BaBar~\cite{BaBarbsg} has presented a preliminary result
from a fully inclusive analysis in which they use lepton tags 
from the other $B$ to almost completely suppress the continuum background (Figure 4).
Note that the lepton tags do not help suppress B decay backgrounds.

\begin{figure}[hbtp]
\hbox to\hsize{\hss
\includegraphics[width=0.8\hsize]{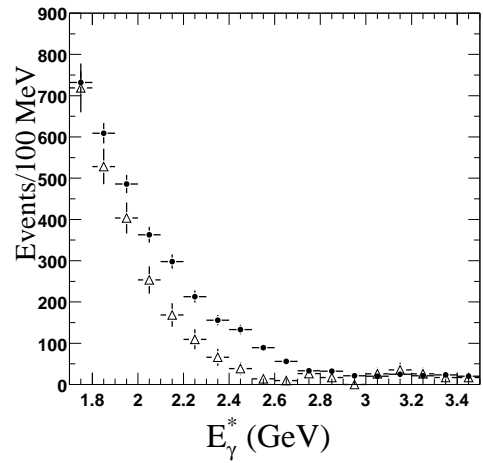}
\hss}
\caption{Preliminary BaBar result for lepton-tagged inclusive $b\to s\gamma$.
Full circles are data, open triangles background (mostly from $b\to c$ decays).}
\label{fig:inclusive2}
\end{figure}

A ``semi-inclusive'' method, which measures a sum of exclusive $B\to X_s\gamma$ 
decays, has been used by both BaBar~\cite{BaBarbxs} and BELLE~\cite{BELLEbxs}.
The hadronic $X_s$ system is reconstructed by BaBar(BELLE) in 12(16) final states
with a mass range up to 2.40(2.05)GeV. This includes about 50\% of all
$b\to s\gamma$ final states. Continuum and B decay backgrounds are subtracted 
by a fit to the beam-constrained B mass in the same way as in an exclusive analysis.   
Results can be shown in terms of the recoil mass $M(X_s)$ (see Figure 5), 
or equivalently in terms of the gamma energy, $E_{\gamma}$, measured in the $B$ 
rest frame. 

\begin{figure}[hbtp]
\hbox to\hsize{\hss
\includegraphics[width=1.0\hsize]{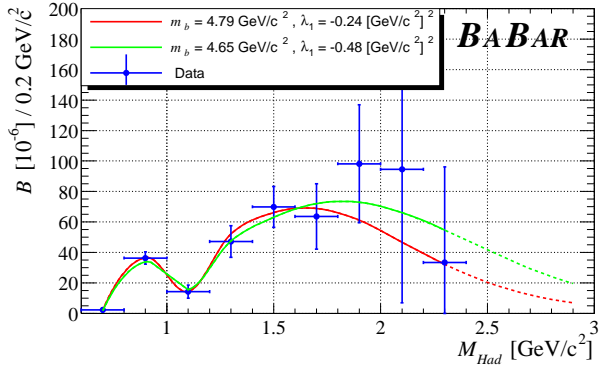}
\hss}
\caption{Preliminary BaBar measurement of hadronic recoil mass in semi-inclusive $b\to s\gamma$}
\label{fig:inclusive10}
\end{figure}

Figure 6 summarizes the measurements of the $b\to s\gamma$ branching fraction.
Computing a world average is complicated by the correlated 
systematic and theoretical errors. 
The fully inclusive method has a dominant systematic error from the B decay 
background subtraction. This is common to the BaBar and CLEO measurements. 
The semi-inclusive method has a dominant systematic error from the 
efficiency for reconstructing the final states, including a correction for 
missing final states that are not considered. Again this is common 
to the BaBar and BELLE measurements. Note that Figure 5 shows that this 
efficiency systematic depends on the theoretical modelling of the spectral 
shape.  

\begin{figure}[hbtp]
\hbox to\hsize{\hss
\includegraphics[width=1.0\hsize]{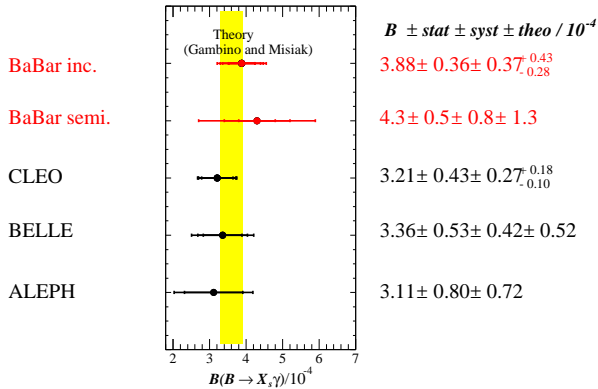}
\hss}
\caption{Summary of $b\to s\gamma$ branching fraction. The shaded band 
shows a recent theoretical calculation}
\label{fig:inclusive3}
\end{figure}

The theoretical error is quoted as the extrapolation of the inclusive rate 
from the measured energy range to the full photon spectrum~\cite{extrabsg}.  
CLEO has a lower threshold (2.0GeV), than BaBar(2.1GeV) and BELLE (2.25GeV). 
Thus the small 7\% theoretical error can only be applied to the CLEO result. 
In taking a world average I have assumed a ``typical'' extrapolation error 
of 10\%. Together with a conservative treatment of the correlated 
experimental systematics, I obtain:
\begin{displaymath}
BF(b\to s\gamma) = (3.47\pm 0.23\pm 0.32\pm 0.35)\times 10^{-4}
\end{displaymath}
For the convenience of theorists setting constraints on new physics
I also give the 90\% C.L. limits:
\begin{displaymath}
2.8\times 10^{-4} < BF(b\to s\gamma) < 4.2\times 10^{-4} 
\end{displaymath}

It is also interesting to look for inclusive CP asymmetries. 
CLEO has published a measurement 
of the asymmetry in the sum of $b\to s\gamma$ and $b\to d\gamma$
from their fully inclusive analysis~\cite{CLEObsgCP}, 
$A_{CP}(b\to (s+d)\gamma) = -0.079\pm0.108\pm 0.022$. 
This asymmetry is exactly zero in the Standard Model,  
and in many extensions, so it is more interesting 
to measure the asymmetry in $b\to s\gamma$ alone. 
This can be done using the semi-inclusive method.  

\section{$B\to\rho\gamma$ and $V_{td}$}

The measurement of the Cabibbo suppressed transition $b\to d\gamma$ 
is difficult because the signal is $\approx 20\times$ smaller, and the 
continuum background is $\approx 3\times$ larger. 
It is also important to remove $b\to s\gamma$ events which are 
a serious background. 

\begin{figure}[hbtp]
\hbox to\hsize{\hss
\includegraphics[width=0.8\hsize]{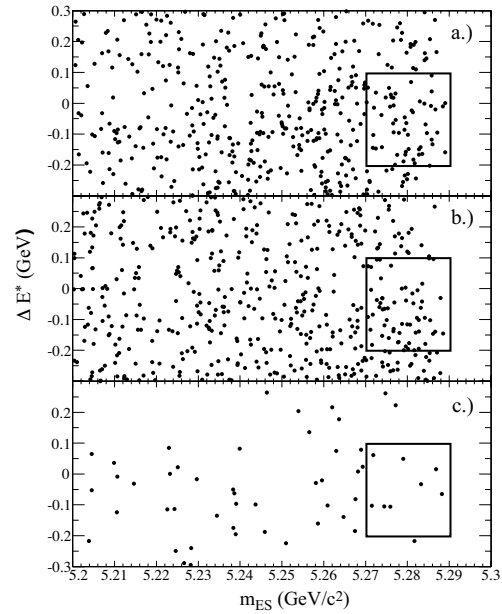}
\hss}
\caption{The BaBar search for $B\to \rho (\omega )\gamma$ with 80fb$^{-1}$ 
of data: (a) $\rho^0\gamma$ (b) $\rho^+\gamma$ (c) $\omega\gamma$}
\label{fig:inclusive4}
\end{figure}

The best limits on the exclusive decays $B\to \rho(\omega)\gamma$ 
come from BaBar~\cite{rhogamma}, which uses a neural network to 
suppress most of the continuum background. The $B\to K^*\gamma$ events 
are removed using particle identification to veto kaons, with a $K\to\pi$ 
fake rate of $\approx$1\%. A multi-dimensional likelihood fit is made 
to the remaining events (Figure 7), to give 90\% C.L. upper limits 
of 1.2, 2.1 and 1.0$\times 10^{-6}$ on $\rho^0\gamma$, $\rho^+\gamma$ 
and $\omega\gamma$ respectively. Assuming isospin symmetry, this gives a 
combined limit $BF(B\to\rho\gamma)<1.9\times 10^{-6}$ (90\% C.L.).

The ratio of $B\to\rho\gamma$ to $B\to K^*\gamma$ can be described by
~\cite{Alirhogamma}:
\begin{displaymath}
{{B\to\rho\gamma}\over{B\to K^*\gamma}} = 
|{{V_{td}}\over{V_{ts}}}|^2
\left({{1-m_{\rho}^2/M_B^2}\over{1-m_{K^*}^2/M_B^2}}\right)^3
\zeta^2[1+\Delta R]
\end{displaymath}
where $\zeta = 0.76\pm 0.10$ is an SU(3) breaking factor
in the ratio of form factors.
$\Delta R=0.0\pm 0.2$ accounts for possible weak annihilation 
and long distance contributions which appear mainly in $B^+\to\rho^+\gamma$.
Eventually these can be checked by comparing $\rho^+\gamma$ and $\rho^0\gamma$.  
Using the above equation gives
$|V_{td}/V_{ts}| < 0.34$.
This constraint on the CKM element $V_{td}$ is shown in the $\rho/\eta$ plane
in Figure 8. It is not as tight as the constraint from $B_s/B_d$ mixing. 
However, new physics may appear in different ways in penguin and mixing 
diagrams, so it is important to measure $V_{td}$ in both processes. 

\begin{figure}[hbtp]
\hbox to\hsize{\hss
\includegraphics[width=0.8\hsize]{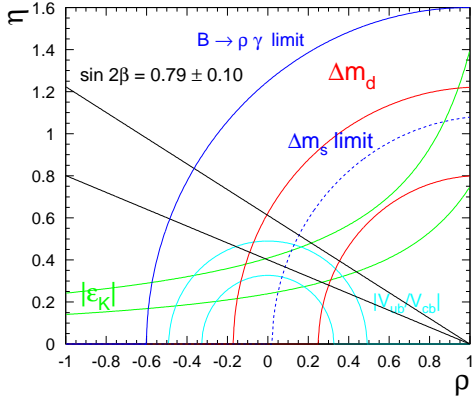}
\hss}
\caption{Constraint on CKM unitarity triangle obtained from 
upper limit on $B\to\rho\gamma$}
\label{fig:inclusive5}
\end{figure}

BaBar and BELLE each expect to collect $\approx 500fb^{-1}$ by 2005.
This should be sufficient to observe $B\to\rho\gamma$. 
It may also be feasible to measure the inclusive $b\to d\gamma$ rate 
using the semi-inclusive method. For the extraction of 
$|V_{td}/V_{ts}|$ the ratio of $b\to d\gamma$ 
to $b\to s\gamma$ has much smaller theoretical uncertainties 
than the ratio of the exclusive decays~\cite{Soni}.

\section{Measurements of $b\to s\ell^+\ell^-$}

These FCNC decays are described by a combination 
of radiative and electroweak penguins, and a box diagram
(Figure 9). This makes them more sensitive to new physics 
contributions than $b\to s\gamma$. There is particular interest in an eventual 
measurement of the forward-backward lepton asymmetry as 
a function of the dilepton mass, $\hat{s} = m_{ll}^2/m_b^2$~\cite{sllafb}.

\begin{figure}[hbtp]
\hbox to\hsize{\hss
\includegraphics[width=0.8\hsize]{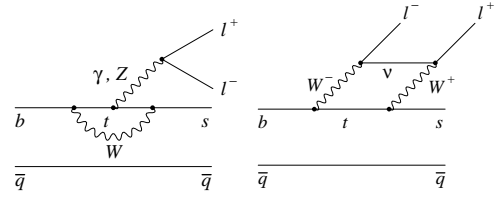}
\hss}
\caption{Diagrams contributing to $b\to s\ell^+\ell^-$}
\label{fig:inclusive6}
\end{figure}

The exclusive decays $B\to K\ell^+\ell^-$, $B\to K^* \mu^+\mu^-$ 
and $B\to K^* e^+e^-$,  are predicted to have branching 
fractions of $(0.35\pm 0.12)$, $(1.19\pm 0.39)$  and $(1.58\pm 0.49)\times 10^{-6}$
respectively~\cite{Alisll}. 
The difference between the $K^*$ decays is due to a pole at low
$\hat{s}$ from the radiative penguin diagram.  

The decay $B\to K\ell^+\ell^-$ was first observed by BELLE~\cite{BELLEkll}, 
and has now been confirmed by BaBar~\cite{BaBarkll}, who also see evidence 
for $B\to K^*\ell^+\ell^-$ (Figure 10). 

\begin{figure}[hbtp]
\hbox to\hsize{\hss
\includegraphics[width=0.8\hsize]{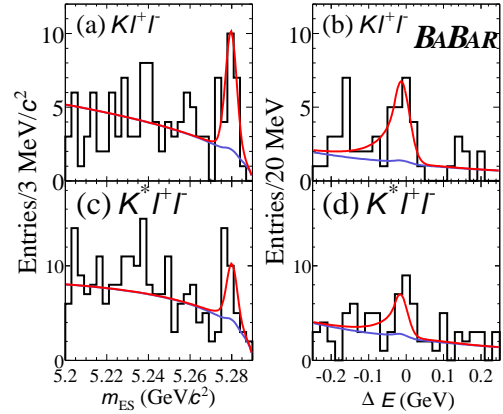}
\hss}
\caption{BaBar measurement of $B\to K\ell^+\ell^-$ 
and preliminary evidence for $B\to K^*\ell^+\ell^-$ using 80fb$^{-1}$ of data}
\label{fig:inclusive7}
\end{figure}

\begin{table}[htbp]
\begin{center}
\begin{tabular}{|l|c|c|} \hline 
          & $B\to K\ell^+\ell^-/10^{-6}$ & $B\to K^*\ell^+\ell^-/10^{-6}$ \\ \hline 
BABAR &$0.78^{+0.24}_{-0.20}\pm 0.15$ & $1.26^{+0.51}_{-0.44}\pm 0.17$\\ 
BELLE &$0.58^{+0.17}_{-0.15}\pm 0.06$ & 
$1.0^{+0.6}_{-0.5}\pm 0.2$\\ \hline
Average   & $0.66^{+0.15}_{-0.13}\pm 0.06$ & 
$1.10^{+0.42}_{-0.35}\pm 0.13$\\ \hline
\end{tabular}
\end{center}
\caption{Summary of measurement of $B\to K^{(*)}\ell^+\ell^-$. The $K^*$ numbers
are quoted in terms of $K^*\mu^+\mu^-$ assuming the theoretical ratio 
$K^*\mu^+\mu^- = 0.75 K^*e^+e^-$.} 
\end{table}

The inclusive rate is predicted to 
be $4.2\pm 0.7(6.9\pm 1.0)\times 10^{-6}$ for $s\mu^+\mu^- (se^+e^-)$.
BELLE has measured this inclusive rate~\cite{BELLEsll} using a semi-inclusive sum of 
final states (Figure 11). With the requirements $m_{ll}>0.2$GeV and $M(X_s)< 2.1$GeV,
they measure
\begin{displaymath}
BF(b\to s\ell^+\ell^-) = 6.1\pm 1.4(stat.)^{+1.4}_{-1.1}(syst.)\times 10^{-6}
\end{displaymath}
which is somewhat larger than the $s\mu^+\mu^-$ prediction.

\begin{figure}[hbtp]
\hbox to\hsize{\hss
\includegraphics[width=0.8\hsize]{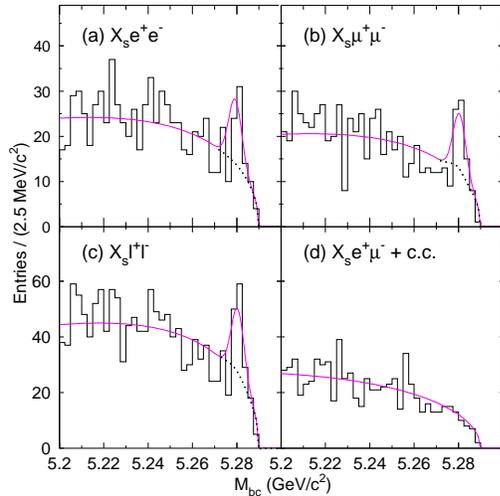}
\hss}
\caption{BELLE measurement of inclusive $b\to s\ell^+\ell^-$}
\label{fig:inclusive8}
\end{figure}

These results already provide interesting new constraints 
on extensions of the Standard Model. With larger data samples
at the B-factories, and eventually at the LHC,  
it will be possible to make precise tests of the theoretical 
predictions for these decays.

\end{document}